\documentclass[pre,floatfix,twocolumn,showpacs]{revtex4}

\usepackage{psfig}

\renewcommand{\vec}[1]{\textbf{#1}}
\def\qq{\qquad\qquad}                      
\def\qqq{\qquad\qquad\qquad}               
\def\q{\qquad}
\def\beg{\begin{eqnarray}}
\def\ende{\end{eqnarray}}

\begin{document}

\title{Theory of current instability experiments in magnetic  
Taylor-Couette flows}
\author{G\"unther R\"udiger, Manfred Schultz}
\affiliation{Astrophysikalisches Institut Potsdam,
         An der Sternwarte 16, D-14482 Potsdam, Germany}
\email{gruediger@aip.de, mschultz@aip.de}
\author{Dima Shalybkov}      
\affiliation{A.F. Ioffe Institute for Physics and Technology,
          194021, St. Petersburg, Russia}
\email{dasha@astro.ioffe.ru} 
\author{Rainer Hollerbach}      
\affiliation{Department of Applied Mathematics, University of 
Leeds, Leeds, LS2 9JT, UK}
\email{rh@maths.leeds.ac.uk}

\date{\today}

\begin{abstract}
We consider the linear stability of dissipative MHD Taylor-Couette
flow with imposed toroidal magnetic fields.  The inner and outer
cylinders can be either insulating or conducting; the inner one rotates,
the outer one is stationary.  The magnetic Prandtl number can be as small
as $10^{-5}$, approaching realistic liquid-metal values.  The magnetic field
destabilizes the flow, except for radial profiles of $B_\phi(R)$ close to
the current-free solution.
The profile with $B_{\rm in}=B_{\rm out}$ (the most uniform field) is
considered in detail. For weak fields the TC-flow is {\em stabilized},
until for moderately strong fields the $m=1$ azimuthal mode dramatically
destabilizes the flow again.  There is thus a maximum value for the
critical Reynolds number.  For sufficiently strong fields (as measured by
the Hartmann number) the toroidal field is always unstable, even for $\rm Re=0$.

The electric currents needed to generate the required toroidal fields in
laboratory experiments are a few kA if liquid sodium is used, somewhat more
if gallium is used.  Weaker currents are needed for wider gaps, so a wide-gap
apparatus could succeed even with gallium.  The critical Reynolds numbers
are only somewhat larger than the nonmagnetic values, so such an experiment
would require only modest rotation rates.
\end{abstract}


\maketitle

\section{Motivation}

Taylor-Couette flows of electrically conducting fluid between rotating
concentric cylinders are a classical problem of hydrodynamic and
hydromagnetic stability theory.  It is becoming increasingly clear that the
stability of differential rotation combined with magnetic fields is also
one of the key problems in MHD astrophysics.  For uniform axial magnetic
fields this phenomenon is called the magnetorotational instability (MRI).
The Keplerian rotation of accretion disks with weak vertical magnetic
fields becomes unstable, so that angular momentum is transported outward
(star formation) and gravitational energy is efficiently transformed into
heat and radiation (quasars). Galactic rotation profiles (approximately
$\Omega\propto 1/R$) may also be MRI-unstable, resulting in interstellar
MHD turbulence \cite{SB99, DER04}.

Due to the small size of laboratory experiments, and the extremely small
magnetic Prandtl numbers of liquid metals, it is very difficult to achieve
the MRI in the lab with a purely axial magnetic field.  However, by adding a
toroidal field $B_\phi\propto 1/R$ (which is current-free within the fluid),
it was possible to obtain the MRI experimentally \cite{S06, R06}, with the
most unstable mode being axisymmetric and oscillatory, just as predicted
theoretically \cite{H05}.

If the field strength is sufficiently great, such a current-free toroidal
field yields a magnetorotational instability even without an axial field;
the instability is simply non-axisymmetric rather than axisymmetric.  The
simultaneous occurrence of a stable differential rotation and a stable
toroidal field may therefore nevertheless be unstable, provided that the
magnetic Reynolds number exceeds $O(100)$ \cite{H06}.

In contrast to these magnetorotational instabilities, in which the magnetic
field acts as a catalyst, but not as a source of energy, toroidal fields
that are not current-free may become unstable more directly, by so-called
current or Tayler instabilities \cite{T73, S99}.  Because the source of
energy is now the current rather than the differential rotation, these
(non-axisymmetric) instabilities can exist even without any differential
rotation, provided only the current is large enough.  The magnitude depends
quite strongly on the precise radial profile of the associated magnetic
field $B_\phi(R)$, but not on the magnetic Prandtl number.  The topic of
this paper is how Tayler instabilities interact with differential rotation,
and whether it might be possible to realize some of the resulting modes in
laboratory experiments.  The combination of Tayler instabilities and
differential rotation may also be relevant to a broad range of astrophysical
problems, including the stability of the solar tachocline, the existence of
active solar longitudes \cite{KS02}, the flip-flop phenomenon of stellar
activity \cite{K01}, A-star magnetism \cite{BN06}, and even the possibility
of `a differential rotation driven dynamo in a stably stratified star'
\cite{B06}.

\section{Equations}

According to the Rayleigh criterion an ideal flow is stable against
axisymmetric perturbations whenever the specific angular momentum increases
outwards
\beg
\frac{{\rm{d}}}{{\rm{d}}R}(R^2\Omega)^2 > 0,
\label{ray}
\ende
where ($R$, $\phi$, $z$) are cylindrical coordinates, and $\Omega$ is the
angular velocity.   The necessary and sufficient condition for the
axisymmetric stability of an ideal Taylor-Couette flow with an imposed
azimuthal magnetic field $B_\phi$ is
\beg
\frac{1}{R^3}\frac{{\rm{d}}}{{\rm{d}}R}(R^2\Omega)^2 - \frac{R}{\mu_0 \rho}
\frac{{\rm{d}}}{{\rm{d}}R}\left( \frac{B_\phi}{R} \right)^2 > 0,
\label{mich}
\ende
where $\mu_0$ is the permeability and $\rho$ the density \cite{M54, C61}.
In particular, {\em all} ideal flows can thus be destabilized, by azimuthal
magnetic fields with the right profiles and amplitudes.  Any fields
increasing outward more slowly than $B_\phi\propto R$, including in particular
the outwardly decreasing current-free field $B_\phi\propto 1/R$, have a
stabilizing influence though \cite{V59}.

Tayler \cite{T73} found the necessary and sufficient condition
\beg
- \frac{{\rm{d}}}{{\rm{d}}R}( R B_\phi^2) > 0.
\label{tay}
\ende
for the non-axisymmetric stability of an ideal fluid at rest.  Outwardly
increasing fields are therefore unstable now (but $B_\phi\propto 1/R$ is
still stable).  If this condition (\ref{tay}) is violated, the most unstable
mode has azimuthal wavenumber $m=1$.  In this paper we wish to consider
how these Tayler instabilities are modified if the fluid is not at rest, but
is instead differentially rotating.

We will find that, depending on the magnitudes of the imposed differential
rotation and magnetic fields, and also on the magnetic Prandtl number, a
magnetic field may either stabilize or destabilize the differential rotation,
and the most unstable mode may be either the axisymmetric Taylor vortex flow,
or the non-axisymmetric Tayler instability.  We focus on the limit of small
magnetic Prandtl numbers appropriate for liquid metals, and calculate the
rotation rates and electric currents that would be required to obtain some
of these instabilities in liquid metal laboratory experiments.

The governing equations are
\begin{eqnarray}
\frac{\partial \vec{U}}{\partial t} + (\vec{U} \nabla)\vec{U} =
-\frac{1}{\rho} \nabla P + \nu \Delta \vec{U} + 
\frac{1}{\mu_0}{\textrm{curl}}\ \vec{B} \times \vec{B},
\label{mhd}
\end{eqnarray}
\begin{eqnarray}
\frac{\partial \vec{B}}{\partial t}= {\textrm{curl}} (\vec{U} \times \vec{B})+ \eta \Delta\vec{B},
\label{mhd1}
\end{eqnarray}
and
\beg
{\textrm{div}}\ \vec{U} = {\textrm{div}}\ \vec{B} = 0,
\label{mhd2}
\ende
where $\vec{U}$ is the velocity, $\vec{B}$ the magnetic field, $P$  the 
pressure, $\nu$ the kinematic viscosity, and $\eta$ the magnetic diffusivity.

\begin{figure}[ht]
\psfig{figure=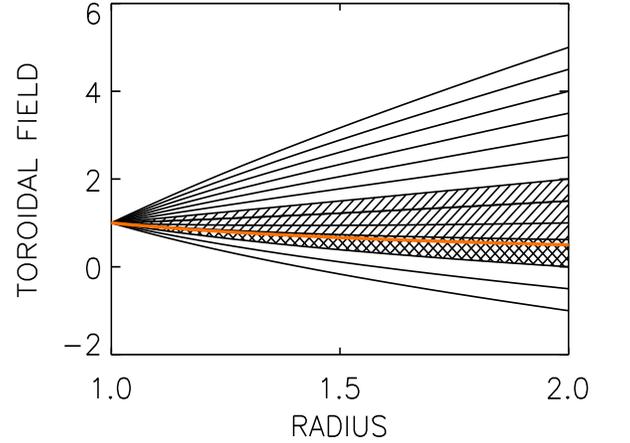,width=9cm,height=7cm}
\caption{\label{fig1} The basic state profiles of $B_\phi(R)$, for $\hat\eta
=0.5$.  Remembering that $B_\phi(1)$ has been normalized to 1, we find that
$\hat\mu_B$ is given simply by $B_\phi(2)$, and can therefore be read off the
right-hand axis.  The stability domain for $m=1$ (see (\ref{m1})) is
cross-hatched; the cross-hatched and hatched domains together are the
stability domain for $m=0$ (see (\ref{m0})). The current-free solution
$B_\phi=1/R$ is given by the gray line ($\hat\mu_B=0.5$).  The electric
currents inside and outside the inner cylinder are parallel above
$\hat\mu_B=0.5$ and anti-parallel below $\hat\mu_B=0.5$, that is, the signs
of $a_B$ and $b_B$ are the same above 0.5, and opposite below.}
\end{figure}

The  basic state is $U_R=U_z=B_R=B_z=0$ and
\beg
U_\phi=R\Omega=a_\Omega R+\frac{b_\Omega}{R},  \q
B_\phi=a_B R+\frac{b_B}{R},
\label{basic}
\ende
where $a_\Omega$, $b_\Omega$, $a_B$ and $b_B$ are constants defined by 
\beg
a_\Omega=\Omega_{\rm{in}}\frac{\hat \mu_\Omega-{\hat\eta}^2}{1-{\hat\eta}^2}, \q
b_\Omega=\Omega_{\rm{in}} R_{\rm{in}}^2 \frac{1-\hat\mu_\Omega}{1-{\hat\eta}^2},
\nonumber \\
a_B=\frac{B_{\rm{in}}}{R _{\rm{in}}}\frac{\hat \eta
(\hat \mu_B - \hat \eta)}{1- \hat \eta^2},  \q
b_B=B_{\rm{in}}R _{\rm{in}}\frac{1-\hat\mu_B \hat\eta}
{1-\hat \eta^2},
\label{ab}
\ende
where
\begin{equation}
\hat\eta=\frac{R_{\rm{in}}}{R_{\rm{out}}}, \; \; \;
\hat\mu_\Omega=\frac{\Omega_{\rm{out}}}{\Omega_{\rm{in}}},  \; \; \;
\hat\mu_B=\frac{B_{\rm{out}}}{B_{\rm{in}}}.
\label{mu}
\end{equation}
$R_{\rm{in}}$ and $R_{\rm{out}}$ are the radii of the inner and outer
cylinders, $\Omega_{\rm{in}}$ and $\Omega_{\rm{out}}$ are their rotation
rates (we will in fact fix $\Omega_{\rm{out}}=0$ for all results presented
here), and $B_{\rm{in}}$ and $B_{\rm{out}}$ are the azimuthal magnetic fields
at the inner and outer cylinders. The possible magnetic field solutions are
plotted in Fig. \ref{fig1}.  Note though that -- unlike $\Omega$, where
$\Omega_{\rm{in}}$ and $\Omega_{\rm{out}}$ are the physically relevant
quantities -- for $B_\phi$ the fundamental quantities are not so much
$B_{\rm{in}}$ and $B_{\rm{out}}$, but rather $a_B$ and $b_B$ themselves.
In particular, a field of the form $b_B/R$ is generated by running an
axial current only through the inner region $R<R_{\rm{in}}$, whereas a
field of the form $a_B R$ is generated by running an axial current through
the entire region $R<R_{\rm{out}}$, including the fluid.  One of the aspects
we will be interested in later on is how large these currents must be, and
whether they could be generated in a laboratory experiment.

We are interested in the linear stability of the basic state (\ref{basic}).
The perturbed quantities of the system are given by
\beg
u_R, \; R\Omega+u_\phi , \; u_z, \; b_R,  \; B_\phi+b_\phi, \; b_z.
\ende
Applying the usual normal mode analysis, we look for solutions of the
linearized equations of the form
\beg
F=F(R){\rm{exp}}\bigl({\rm{i}}(kz+m\phi+\omega t)\bigr).
\label{nmode}
\ende
The dimensionless numbers of the problem are the magnetic Prandtl number
Pm, the Hartmann number Ha, and the Reynolds number Re, given by
\beg
{\rm{Pm}} = \frac{\nu}{\eta}, \q
{\rm{Ha}}=\frac{B_{\rm{in}} R_0}{\sqrt{\mu_0 \rho \nu \eta}},  \q 
{\rm{Re}}=\frac{\Omega_{\rm{in}} R_0^2}{\nu},
\label{pm}
\ende
where $R_0=(R_{\rm{in}}(R_{\rm{out}}-R_{\rm{in}}))^{1/2}$ is the unit of length.

Using (\ref{nmode}), linearizing the equations (\ref{mhd}) and (\ref{mhd1}),
and representing the result as a system of first order equations, we have
\begin{eqnarray}
\lefteqn{\frac{du_R}{dR}+\frac{u_R}{R}+{\textrm i}\frac{m}{R}u_\phi+{\textrm i}ku_z=0,}
\nonumber \\
\lefteqn{\frac{dP}{dR}+{\textrm i}\frac{m}{R}X_2+{\textrm i}kX_3
+\left(k^2+\frac{m^2}{R^2}\right)u_R+}
\nonumber \\
&& \qq \qq +{\textrm{iRe}}(\omega+m\Omega)u_R-2\Omega{\textrm Re} u_\phi- 
\nonumber \\
&& \qq \qq -{\textrm{iHa}}^2\frac{m}{R}B_\phi b_R +2{\textrm{Ha}}^2\frac{B_\phi}{R}
b_\phi=0,
\nonumber \\
\lefteqn{\frac{dX_2}{dR}-\left(k^2+\frac{m^2}{R^2}\right)u_\phi-
{\textrm{iRe}}(\omega+m\Omega)u_\phi+}
\nonumber \\
&& \qq \qq +2{\textrm i}\frac{m}{R^2}u_R
-\frac{{\textrm{Re}}}{R}\frac{d}{dR}\left(R^2\Omega\right)u_R+
\nonumber \\
&&+\frac{{\textrm{Ha}}^2}{R}\frac{d}{dR}\left(R B_\phi\right)b_R
+{\textrm{iHa}}^2\frac{m}{R}B_\phi b_\phi
-{\textrm{i}}\frac{m}{R}P=0,
\nonumber \\
\lefteqn{\frac{dX_3}{dR}+\frac{X_3}{R}-\left(k^2+\frac{m^2}{R^2}\right)u_z-
{\textrm{iRe}}(\omega+m\Omega)u_z-}
\nonumber \\
&& \qqq \qq -{\textrm i}kP+{\textrm{iHa}}^2\frac{m}{R}B_\phi b_z=0,
\nonumber \\
\lefteqn{\frac{db_R}{dR}+\frac{b_R}{R}+{\textrm i}\frac{m}{R}b_\phi+{\textrm i}kb_z=0,}
\nonumber \\
\lefteqn{\frac{db_z}{dR}-\frac{{\textrm i}}{k}\left(k^2+\frac{m^2}{R^2}\right)b_R
+{\textrm{PmRe}}\frac{1}{k}(\omega+m\Omega)b_R+}
\nonumber \\
&& \qqq \qq +\frac{1}{k}\frac{m}{R}X_4-\frac{1}{k}\frac{m}{R}B_\phi u_R=0,
\nonumber \\
\lefteqn{\frac{dX_4}{dR}-\left(k^2+\frac{m^2}{R^2}\right)b_\phi
-{\textrm{iPmRe}}(\omega+m\Omega)b_\phi+}
\nonumber \\
&& \q + {\textrm i}\frac{2m}{R^2}b_R
-R\frac{d}{dR}\left(\frac{B_\phi}{R}\right)u_R
+{\textrm{PmRe}}R\frac{d\Omega}{dR}b_r+
\nonumber \\
&& \qqq \qqq +{\textrm i}\frac{m}{R}B_\phi u_\phi =0,
\label{syst}
\end{eqnarray}
where $X_2$, $X_3$ and $X_4$ are defined as
\begin{equation}
X_2=\frac{du_\phi}{dR}+\frac{u_\phi}{R},
\ \ \ \
X_3=\frac{du_z}{dR}, \ \ \ \ 
X_4=\frac{db_\phi}{dR}+\frac{b_\phi}{R}.
\label{def}
\end{equation}
Length has been scaled by $R_0$, time by
$\Omega_{\rm{in}}^{-1}$, the basic state angular velocity by
$\Omega_{\rm{in}}$, the perturbation velocity by $\eta/R_0$, and the
magnetic fields, both basic state and perturbation, by $B_{\rm{in}}$.

An appropriate set of ten boundary conditions is needed to solve the
system (\ref{syst}).  For the velocity the boundary conditions are always
no-slip,
\beg
u_R=u_\phi=u_z=0.
\label{ubnd}
\ende
For the magnetic field the boundary conditions depend on whether the walls
are insulators or conductors.  For conducting walls the radial component
of the field and the tangential components of the current must vanish,
yielding
\beg
db_\phi/dR + b_\phi/R = b_R = 0.
\label{bcond}
\ende
These boundary conditions are applied at both $R_{\rm{in}}$ and $R_{\rm{out}}$.

For insulating walls the boundary conditions are somewhat more complicated;
matching to interior and exterior potential fields then yields
\beg
b_\phi={m\over kR} b_z,
\ende
\beg
b_R+ {{\textrm i} b_z \over I_m(kR)} \left({m\over kR} I_m(kR)+I_{m+1}(kR)\right)=0,
\label{nonin}
\ende
at $R=R_{\rm{in}}$, and
\beg
b_\phi={m\over kR} b_z,
\ende
\beg
b_R+{{\textrm{i}} b_z \over K_m(kR)} \left({m\over kR} K_m(kR) - K_{m+1}(kR)\right)=0
\label{nonout}
\ende
at $R=R_{\rm{out}}$, where $I_n$ and $K_n$ are the modified Bessel
functions \cite{RS02}.

Given the basic state (\ref{basic}), Tayler's stability condition
(\ref{tay}) to nonaxisymmetric perturbations becomes
\begin{equation}
0<\hat\mu_B<\frac{4\hat\eta(1-\hat\eta^2)}{3-2\hat\eta^2-\hat\eta^4}
\equiv \hat\mu_{1}.
\label{m1}
\end{equation}
Note that $\hat\mu_1\to 1$ (but is always less than 1) if $\hat\eta\to 1$.
For $\hat \eta=0.5$ we have $\hat\mu_1=0.62$.  Similarly, the stability
condition to axisymmetric perturbations becomes
\begin{equation}
0<\hat\mu_B<\frac{1}{\hat\eta} \equiv \hat\mu_{0}.
\label{m0}
\end{equation}
For $\hat \eta=0.5$ we have $\hat\mu_0=2$.
For $0<\hat\eta<1$ we always have $\hat\mu_{1}<\hat\mu_{0}$,
so that the stability interval (\ref{m1}) for $m=1$ is much smaller than
the stability interval (\ref{m0}) for $m=0$, as shown in Fig. \ref{fig1}.
The current-free solution $\hat\mu_B=0.5$ is of course always stable.

\section{Basic Results}
\begin{figure*}[htb]
\vbox{
\hbox{
\psfig{figure=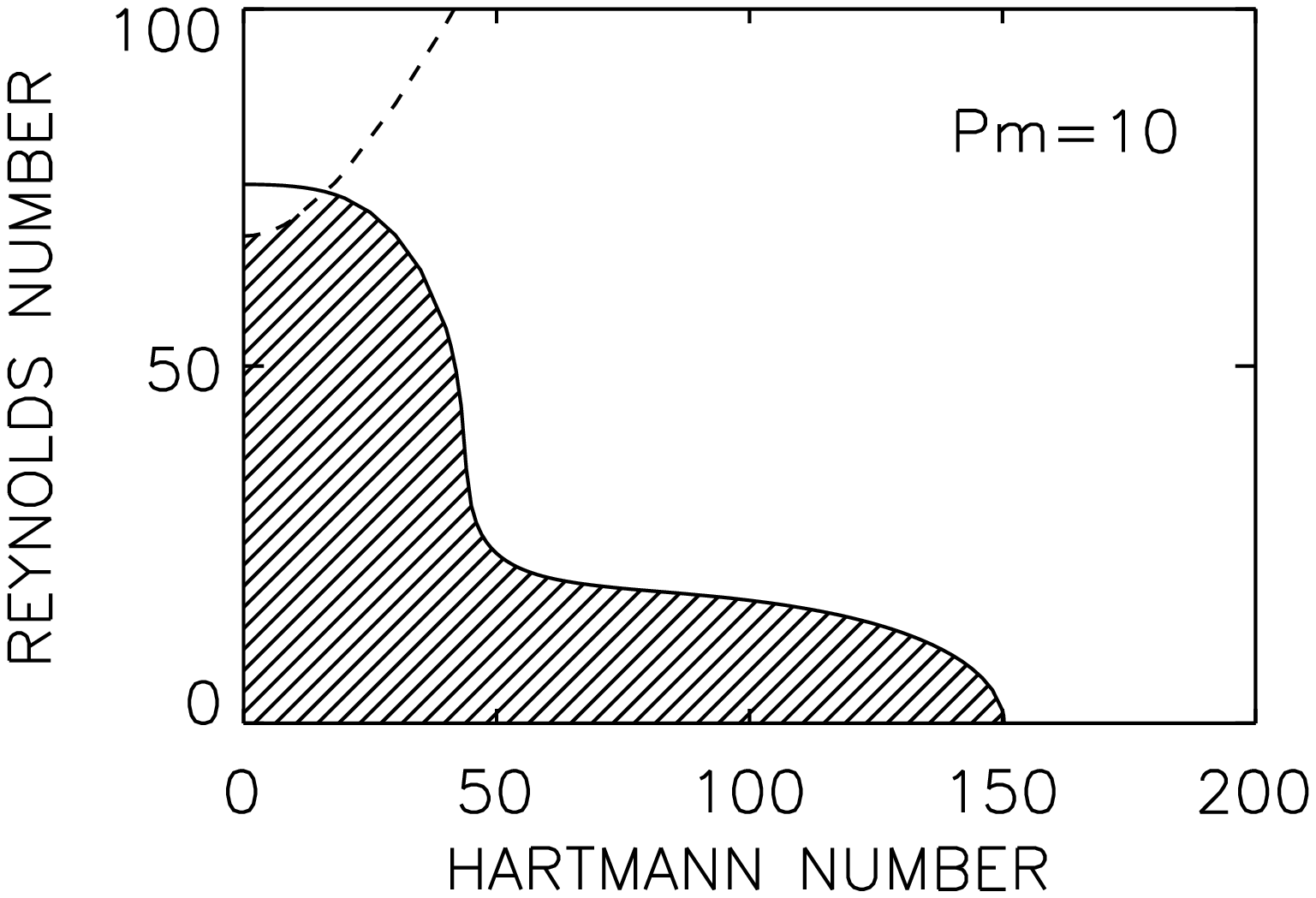,width=8cm,height=5cm}
\psfig{figure=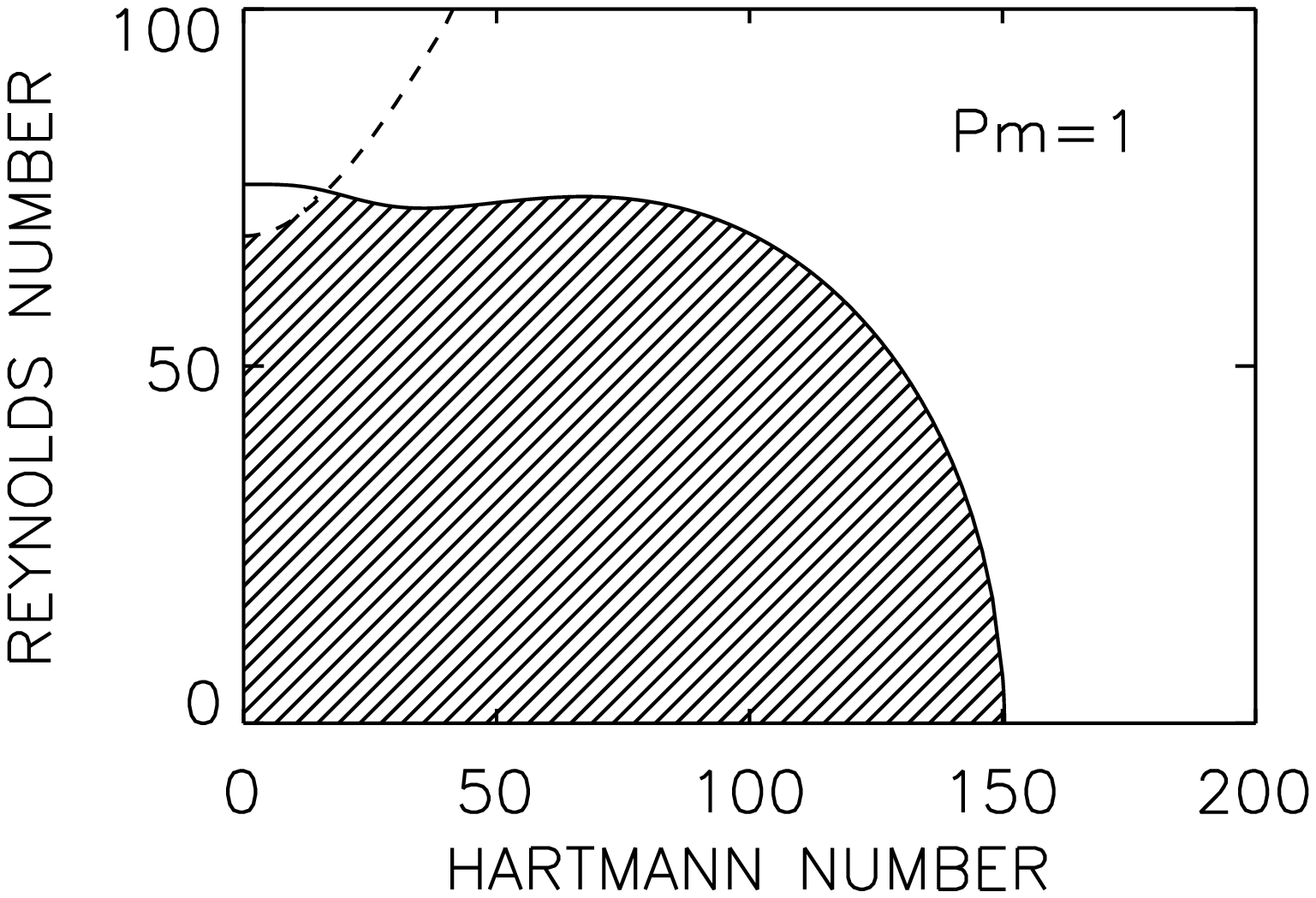,width=8cm,height=5cm}}
\hbox{
\psfig{figure=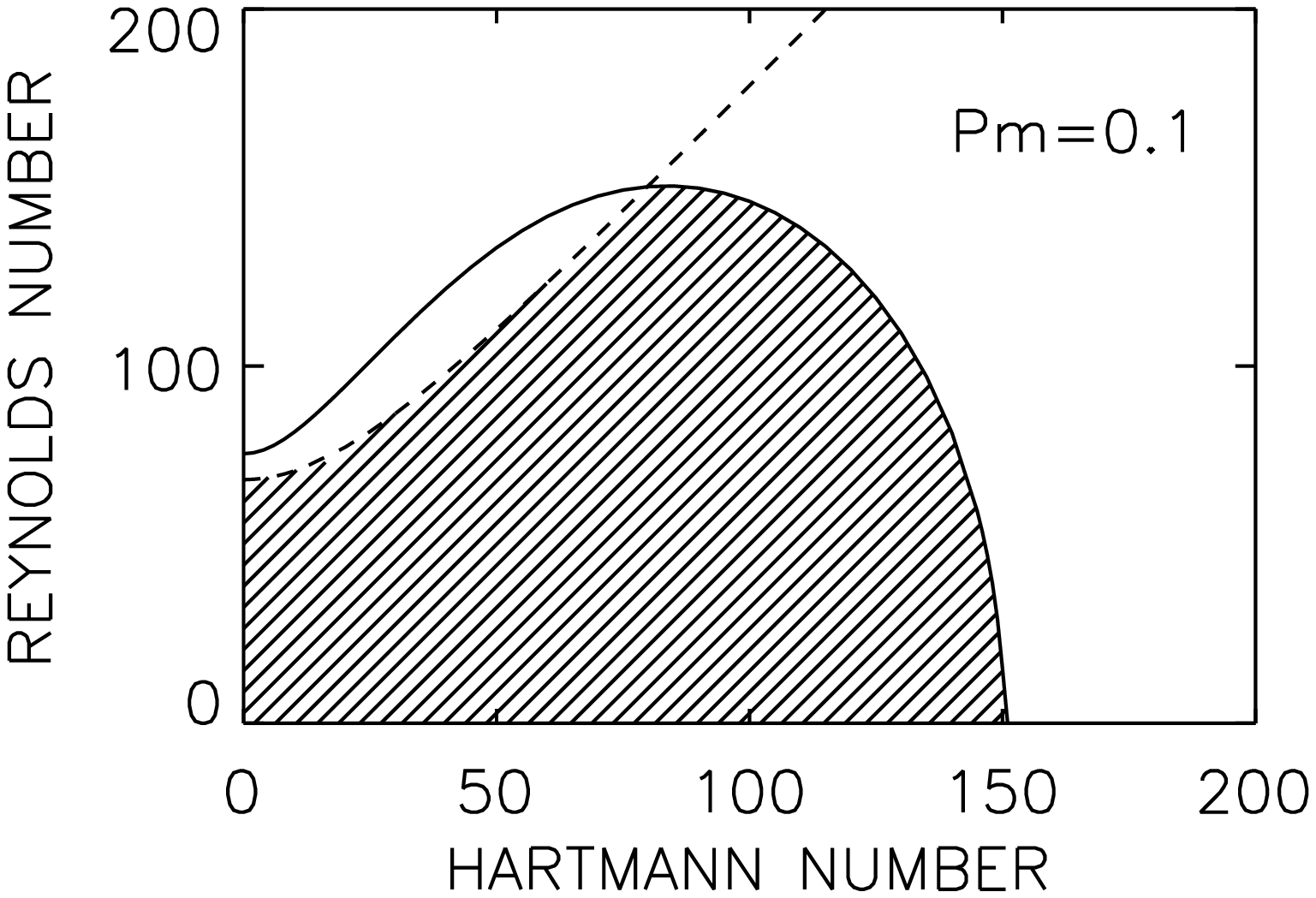,width=8cm,height=5cm}
\psfig{figure=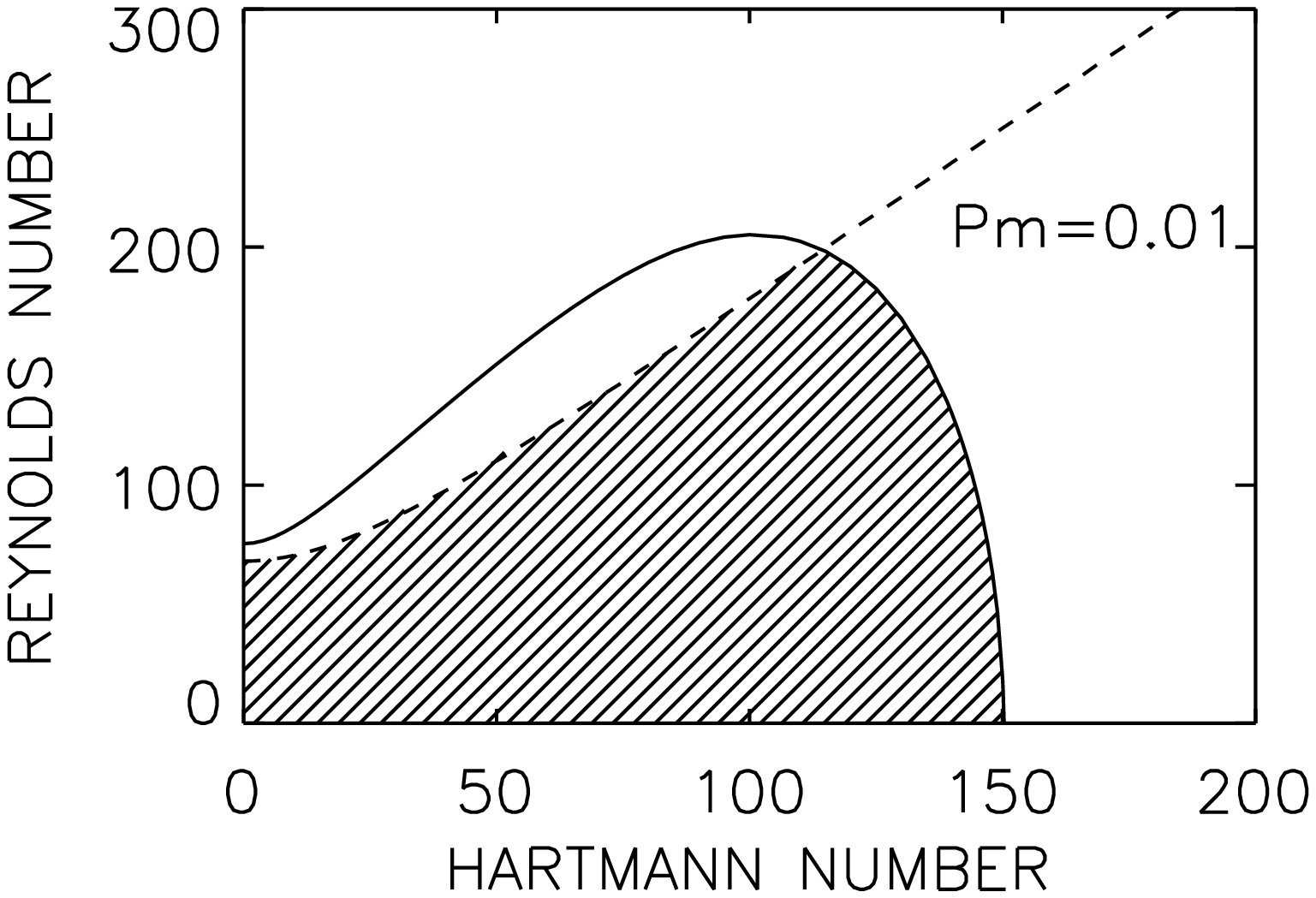,width=8cm,height=5cm}}
}
\caption{\label{pm1} Marginal stability curves for $m=0$ (dashed) and $m=1$
(solid).  The hatched area is thus the region that is stable to both.
Magnetic Prandtl numbers as indicated on each plot.}
\end{figure*}

Figure \ref{pm1} shows how the stability curves depend on $\rm Ha$ and
$\rm Re$, for $\rm Pm=10$, 1, 0.1 and 0.01, with $\hat\mu_\Omega=0$
(stationary outer cylinder) and $\hat\mu_B=1$ ($B_\phi$ as uniform as
possible).  For $\rm Ha=0$ we find that $m=0$ goes unstable before $m=1$,
at the critical Reynolds number $\rm Re_{crit}=68$; this is just the
familiar value for the onset of nonmagnetic Taylor vortices (at this
particular radius ratio).  Being entirely nonmagnetic, this value obviously
does not depend on $\rm Pm$.  At the other limiting case, $\rm Re=0$,
we find that only $m=1$ goes unstable, at the critical Hartmann number
$\rm Ha_{crit}=150$.  These Tayler instabilities also turn out to be
independent of $\rm Pm$, despite being driven by the magnetic field.

We are interested in how these two limiting cases $\rm Ha=0$ and
$\rm Re=0$ are connected, and how the two types of instabilities interact
when neither parameter is zero.  The two instabilities certainly are
connected; the $m=1$ modes in the two limiting cases are smoothly joined
to one another for all  Prandtl numbers.  The nature of the interaction is
quite different though, depending on $\rm Pm$.

Turning to $\rm Pm=1$ first, we see that there is relatively little
interaction between rotational and magnetic effects; instability simply
sets in as long as either $\rm Ha>Ha_{crit}$ or $\rm Re>Re_{crit}$.
For $\rm Pm=10$ the situation is very different.  There we find a broad
range of parameters, for example $\rm Ha=100$ and $\rm Re=50$, that would
be stable if rotational or magnetic effects were acting alone, but which
are now unstable, due to the interaction between the two  (see also
\cite{GF97}).  Finally, for $\rm Pm=0.1$ we have the opposite situation,
namely a range of parameters, for example $\rm Ha=100$ and $\rm Re=100$,
that would be unstable if rotational effects were acting alone, but which
are now stable.  

Small Pm are generally stabilizing. The opposite is true for $\rm Pm>1$.  As shown in Fig. \ref{pm1} (top) instability then also sets in for Hartmann
numbers less than 150.  In other words, for Hartmann numbers exceeding around
50, the critical Reynolds number for the onset of instability is much smaller
than 68.
\section{Liquid metals}
Having explored the general behavior for a range of magnetic Prandtl numbers,
we now focus attention on the limit of very small Pm, such as would apply for
experiments involving liquid metals.  We will here consider conducting and
insulating boundary conditions separately.
\subsection{Conducting cylinder walls}
\begin{figure*}[htb]
\vbox{
\hbox{
\psfig{figure=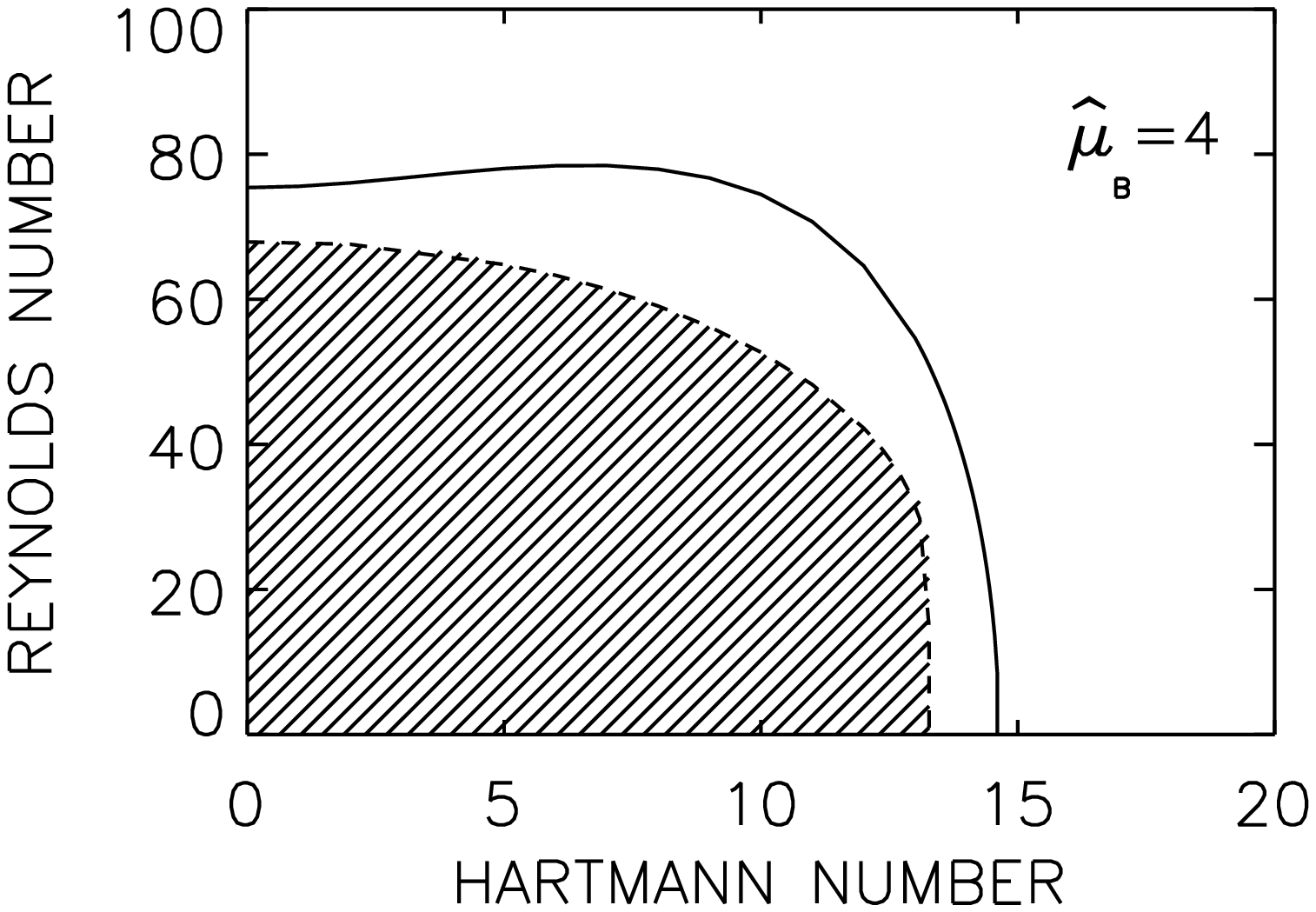,width=6cm,height=5cm}
\psfig{figure=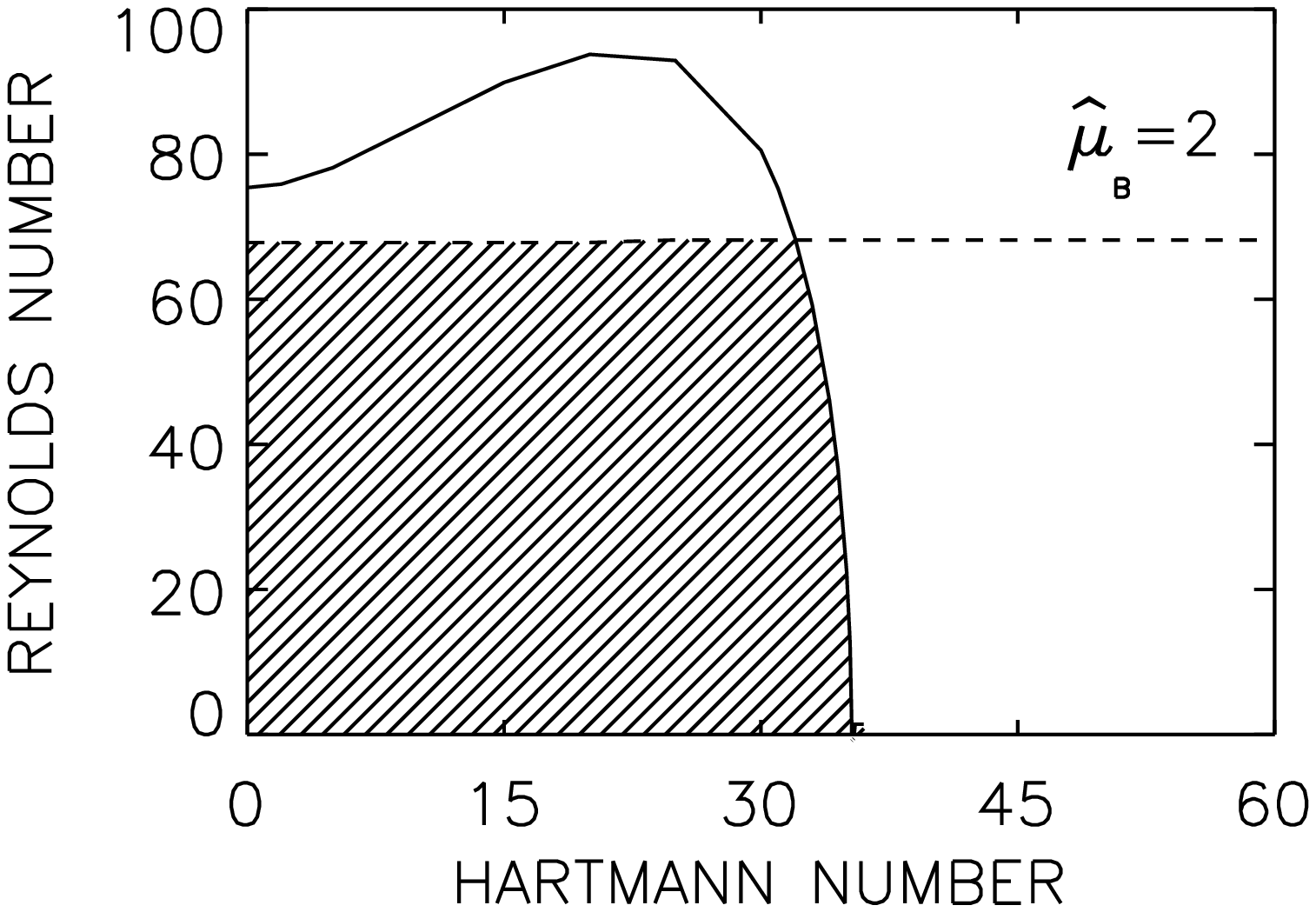,width=6cm,height=5cm}
\psfig{figure=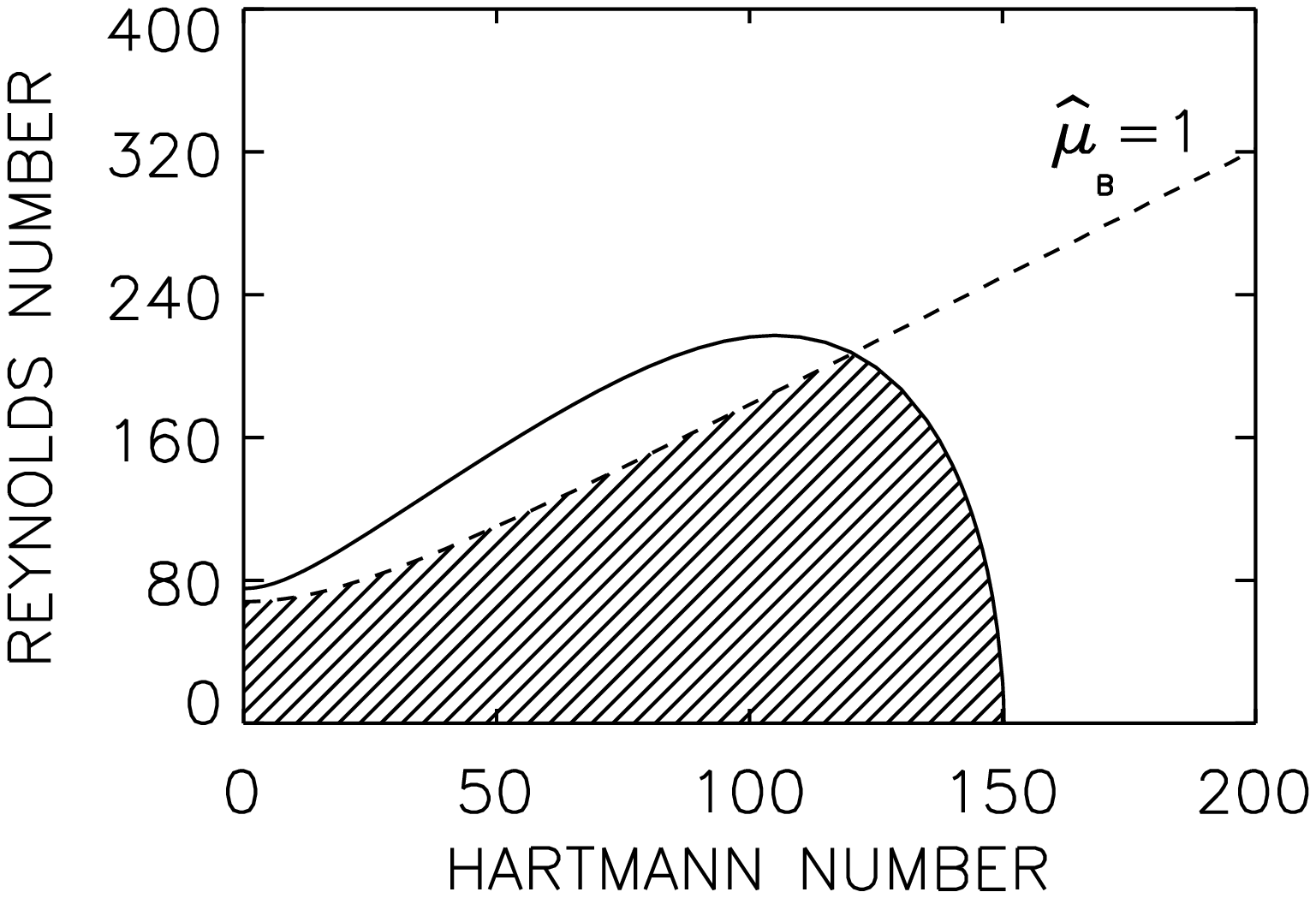,width=6cm,height=5cm}
}
\hbox{
\psfig{figure=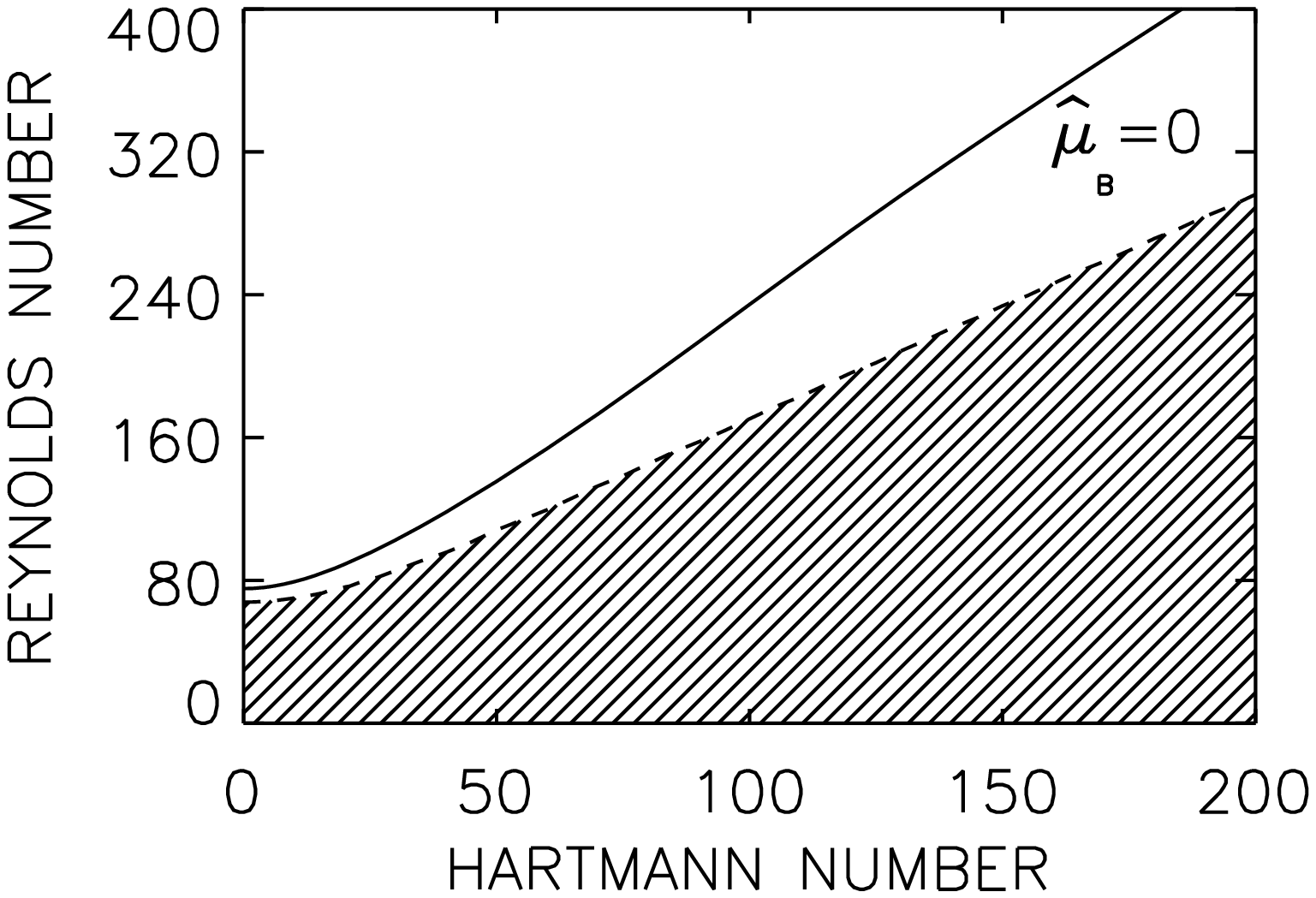,width=6cm,height=5cm}
\psfig{figure=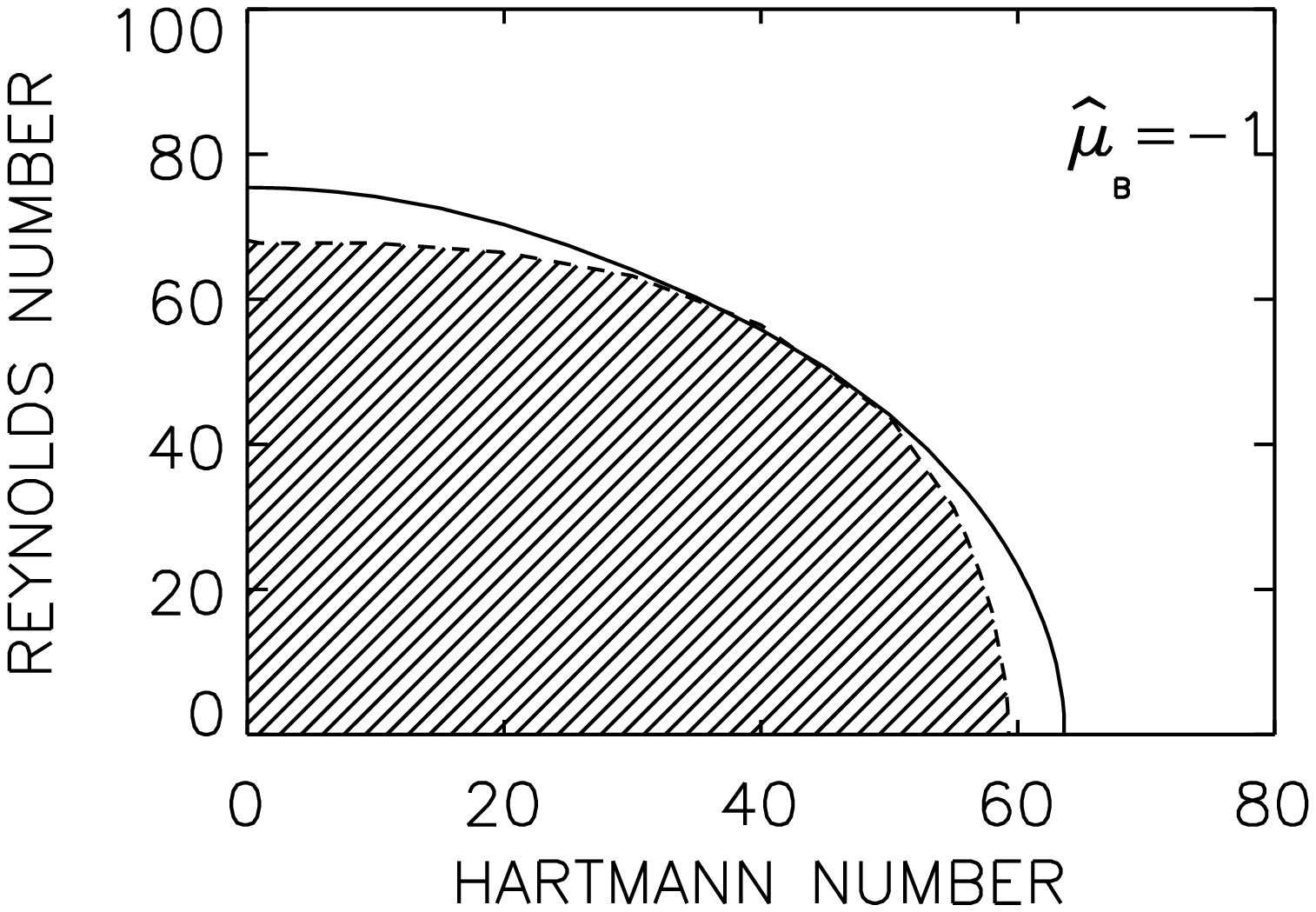,width=6cm,height=5cm}
\psfig{figure=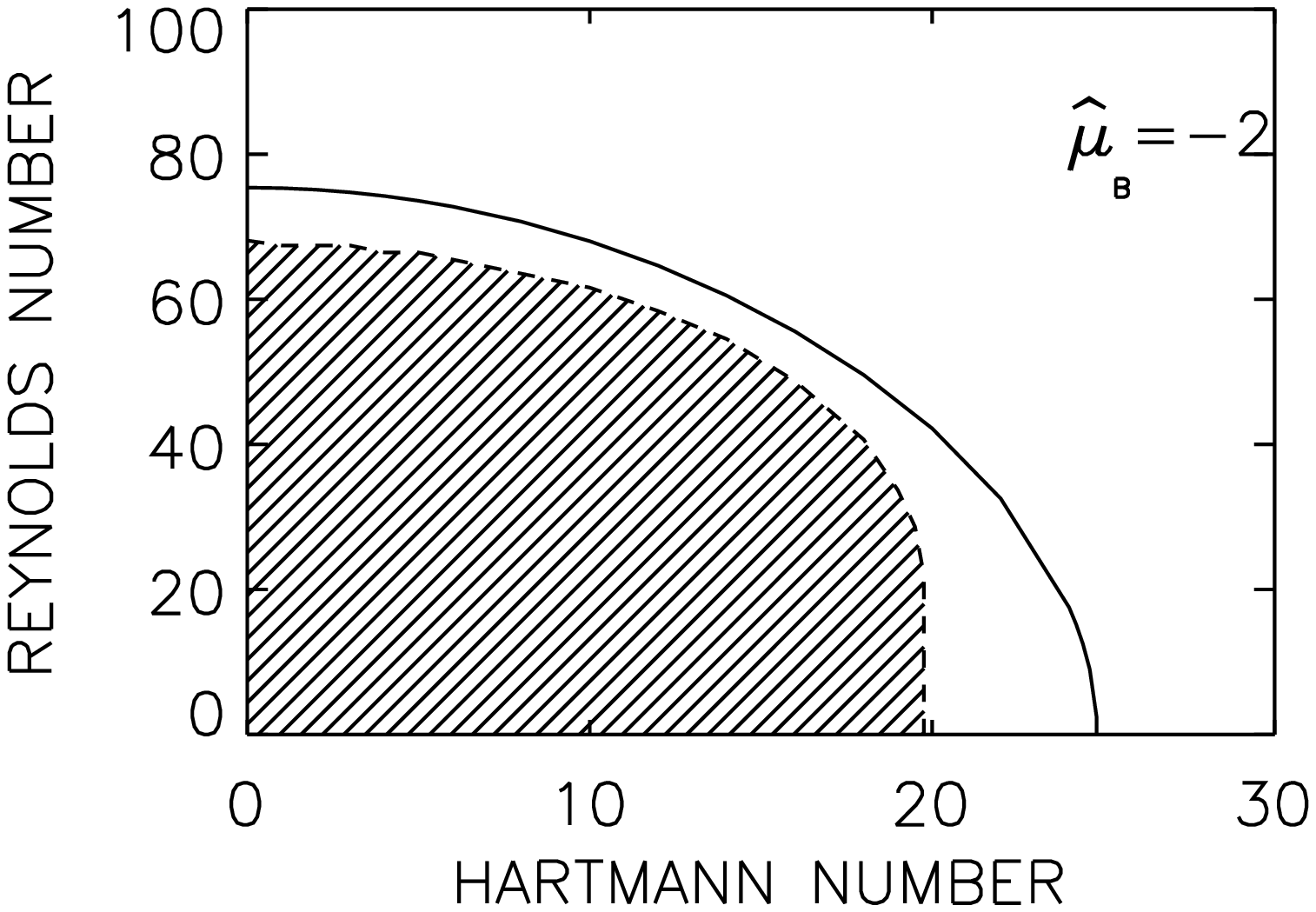,width=6cm,height=5cm}
}
}
\caption{\label{fig2} The marginal stability curves for $m=0$ (dashed)
and $m=1$ (solid).  Pm$=10^{-5}$, $\hat\eta=0.5$, $\hat\mu_\Omega=0$,
and $\hat\mu_B$ as indicated.  Note also how the critical Reynolds numbers
are always of the same order of magnitude as the nonmagnetic result 68,
which is very easy to achieve in the laboratory.}
\end{figure*}

Figure \ref{fig2} shows results for various values of $\hat\mu_B$;
$a_B$ and $b_B$ are the same sign for the values on the left, and the
opposite sign for the values on the right.  The profile that is closest to 
being current-free is $\hat\mu_B=0$, and indeed we find there that even for
Ha=200 there is no sign of any destabilizing influence of the field, for
either axisymmetric or nonaxisymmetric perturbations.  For $0<\hat\mu_B<
\hat\mu_{1}$ the magnetic field stabilizes the flow for both $m=0$ and
$m=1$.

For $\hat\mu_{1}<\hat\mu_B<\hat\mu_{0}$ 
the $m=1$ mode should be unstable, while the $m=0$ mode should be stable.
The values $\hat\mu_B=1$ and $\hat\mu_B=2$ are examples of this situation.
There is always a crossover point at which the most unstable mode changes
from $m=0$ to $m=1$.  Note also how for $\hat\mu_B=1$, the critical Reynolds
number increases for the $m=0$ mode, before suddenly decreasing for the
$m=1$ mode (Fig. \ref{fig2}, left bottom plot).  We have the interesting
situation therefore that weak fields initially stabilize the TC-flow,
before stronger fields eventually destabilize it, via a non-axisymmetric
mode.  Beyond $\rm Ha=150$ (the same value we saw previously in
Fig. \ref{pm1}), the flow is unstable even for $\rm Re=0$.

Except for the almost current-free profile $\hat\mu_B=0$, all other values
share this feature, that there is a critical Hartmann number beyond which
the basic state is unstable even for $\rm Re=0$.  Let Ha$^{(0)}$ and
Ha$^{(1)}$ denote these critical Hartmann numbers, for $m=0$ and 1
respectively.  For the profiles with the largest gradients both modes are
unstable. Strikingly, in these cases $m=0$ is always more unstable than
$m=1$, that is, $\rm Ha^{(0)}<Ha^{(1)}$, see the plots for $\hat\mu_B=4$
and $\hat\mu_B=-2$ of Fig. \ref{fig2}.

\subsection{The required electric currents}
Let $I_{\rm{axis}}$ be the axial current inside the inner cylinder and 
$I_{\rm{fluid}}$  the axial current through  the fluid (i.e. between inner and outer
cylinder). The toroidal field amplitudes at the inner and outer cylinders are
then
\beg
B_{\rm{in}}=\frac{I_{\rm{axis}}}{5R_{\rm{in}}}, \q
B_{\rm{out}}=\frac{(I_{\rm{axis}}+I_{\rm{fluid}})}{5R_{\rm{out}}},
\label{bi}
\ende
where $R$, $B$ and $I$ are measured in cm, Gauss and Ampere. Expressing $I_{\rm{axis}}$ 
and $I_{\rm{fluid}}$ in terms of our dimensionless parameters one finds 
\beg
I_{\rm{axis}}=
5 {\rm Ha}\ \frac{\hat\eta^{1/2}}{(1-\hat\eta)^{1/2}} (\mu_0\rho\nu\eta)^{1/2}
\label{Iin}
\ende
and
\beg
I_{\rm{fluid}}=\frac{\hat\mu_B-\hat\eta}{\hat\eta}I_{\rm{axis}},
\label{Iout}
\ende
in Ampere.  Note also how the required currents depend on the radius {\em
ratio} $\hat\eta$, but not on the actual physical dimensions $R_{\rm{in}}$
and $R_{\rm{out}}$.  Making the entire device bigger thus reduces the
current {\em density}, inversely proportional to the square of the size.
By making the device sufficiently large one can thereby prevent Ohmic
heating within the fluid from becoming excessive.

\begin{table}[h]
\caption{\label{tabmat} Material parameters of liquid metals that might
be used for magnetic TC-experiments.}
\medskip
\begin{tabular}{l|cccc}
\hline
 & $\rho$ [g/cm$^3$] & $\nu$ [cm$^2$/s] & $\eta$ [cm$^2$/s] & $\sqrt{\mu_0 \rho
 \nu \eta}$\\[0.5ex]
\hline
sodium & 0.92 & $7.10\cdot 10^{-3}$ & $0.81\cdot 10^3$ & 8.15\\[0.5ex]
gallium-indium-tin & 6.36 & $3.40\cdot 10^{-3}$ & $2.43\cdot 10^3$ & 25.7\\[0.5ex]
    \hline
 \end{tabular}
 \end{table}

The results for the critical Hartmann numbers are now applied to two different
conducting liquid metals, sodium and gallium-indium-tin \cite{S06}, whose
material parameters are given in Table \ref{tabmat}.  We also wish to consider
the effect of varying the radius ratio $\hat\eta$.  Tables \ref{t1}--\ref{t3}
give the values of the electric currents needed to reach the {\em lesser} of
Ha$^{(0)}$ and Ha$^{(1)}$, for the three values $\hat\eta=0.25$, 0.5 and 0.75,
and for $\hat\mu_B$ ranging from $-10$ to 10 in each case.  Note that for
large $|\hat\mu_B|$, Ha$^{(0)}$ scales as 1/$\hat\mu_B$, and $I_{\rm{fluid}}$
approaches a  constant value.  The calculated currents are lower for fluids
with smaller $\sqrt{\mu_0\rho\nu\eta}$ (i.e. sodium is better than gallium). 

In these Tables, the most interesting experiment, with the almost uniform
field $\hat \mu_B=1$ (see Fig. \ref{fig2}, top-right) is indicated in bold.
For a container with a medium gap of $\hat\eta=0.5$, parallel currents along
the axis and through the fluid of 6.16 kA for sodium and 19.4 kA for gallium
are necessary.  Such sodium experiments should indeed be possible.
Experiments with a wider $\hat\eta=0.25$ gap are even easier; in that case
even gallium experiments should be possible, with a current of 9.29 kA
required (see Table \ref{t1}).

The Reynolds numbers that would be required to obtain not just these $\rm Re=0$
pure Tayler instabilities, but also the transition points from $m=0$ to $m=1$
are also not difficult to achieve; for $R_{\rm{out}}\sim10$ cm, say, rotation
rates of order $10^{-2}$ Hz are already enough.

\begin{table}[h]
\begin{ruledtabular}
\caption{\label{t1} Characteristic Hartmann numbers and electric currents for
a wide gap container ($\hat\eta=0.25$) with conducting walls, using either  
sodium or gallium-indium-tin (in brackets).}
\begin{tabular}{lccccc}
\hline
$\hat\mu_B$ & Ha$^{(0)}$ & Ha$^{(1)}$ & $I_{\rm{axis}}$ [kA] &  $I_{\rm{fluid}}$ [kA]\\
\hline
-10 & 2.29 & 2.05 & 0.0483 (0.152) & -1.98 (-6.24) \\
-5 & 4.23 & 3.98 & 0.0937 (0.296) & -1.97 (-6.21) \\
-4 & 5.14 & 4.93 & 0.116 (0.366) & -1.97 (-6.22)  \\
-3 & 6.63 & 6.50  & 0.153 (0.483) & -1.99 (-6.27)  \\
-2 & 9.69 & 9.71 & 0.228 (0.721) & -2.05 (-6.49) \\
-1 & 24.7 & 22.5 & 0.530 (1.67) & -2.65 (-8.35) \\
\hline
{\bf 1} & $\infty$ & {\bf 41.7} &{\bf 0.982 (3.10)} & {\bf 2.94 (9.29)}  \\
2 & $\infty$ & 13.8 & 0.325 (1.02) & 2.27 (7.17)  \\
3 & $\infty$ & 8.27 & 0.195 (0.614) & 2.14 (6.75) \\
4 & $\infty$ & 5.93 & 0.140 (0.440) & 2.09 (6.60) \\
5 & 10.8 & 4.63 & 0.109 (0.344) & 2.07 (6.53) \\
10 & 3.23 & 2.205 & 0.0519 (0.164) & 2.02 (6.39) \\
\hline
\end{tabular}
\end{ruledtabular}
\end{table}

\begin{table}[h]
\begin{ruledtabular}
\caption{\label{t2} The same as in Table \ref{t1} but for $\hat\eta=0.5$}
\begin{tabular}{lccccc}
\hline
$\hat\mu_B$ & Ha$^{(0)}$ & Ha$^{(1)}$ & $I_{\rm{axis}}$ [kA] &  $I_{\rm{fluid}}$ [kA]\\
\hline
-10 & 3.96 & 5.02 & 0.161 (0.509) & -3.39 (-10.7) \\
-5 & 7.73 & 9.85 & 0.315 (0.994) & -3.47 (-10.9) \\
-4 & 9.61  & 12 & 0.392 (1.24) & -3.53 (-11.1)  \\
-3 & 12.8 & 16.2  & 0.522 (1.65) & -3.65 (-11.5)  \\
-2 & 19.8 & 24.8 & 0.807 (2.55) & -4.04 (-12.7) \\
-1 & 59.3 & 63.7 & 2.42 (7.63) & -7.25 (-22.9) \\
\hline
{\bf 1} & $\infty$ & {\bf 151} & {\bf 6.16 (19.4)} & {\bf 6.16 (19.4)}  \\
 2 & $\infty$ & 35.3 & 1.44 (4.54) & 4.32 (13.6)  \\
3 & 21.0 & 20.6 & 0.840 (2.65) & 4.20 (13.2) \\
4 & 13.2 & 14.6 & 0.538 (1.70) & 3.77 (11.9) \\
5 & 9.84 & 11.4 & 0.401 (1.27) & 3.61 (11.4) \\
10 & 4.44 & 5.4 & 0.181 (0.571) & 3.44 (10.8) \\
\hline
\end{tabular}
\end{ruledtabular}
\end{table}

\begin{table}[h]
\begin{ruledtabular}
\caption{\label{t3} The same as in Table \ref{t1} but for $\hat\eta=0.75$
(narrow gap container).}
\begin{tabular}{lccccc}
\hline
$\hat\mu_B$ & Ha$^{(0)}$ & Ha$^{(1)}$ & $I_{\rm{axis}}$ [kA] &  $I_{\rm{fluid}}$ [kA]\\
\hline
-10 & 9.27 & 12.2 & 0.655 (2.06) & -9.39 (-29.6) \\
-5 & 18.6 & 24.1 & 1.31 (4.13) & -10.0 (-31.5) \\
-4 & 23.3 & 30.1 & 1.65 (5.20) & -10.5 (-33.1)  \\
-3 & 31.6 & 40.3 & 2.23 (7.03) & -11.1 (-35.0)  \\
-2 & 50.4 & 63.4 & 3.56 (11.2) & -13.1 (-41.3) \\
-1 & 163. & 177. & 11.5 (36.3) & -26.8 (-84.5) \\
\hline
{\bf 1} & $\infty$ & {\bf 632} & {\bf 44.6 (141)} & {\bf 14.9 (47.0)}  \\
2 & 66.8 & 87.3 & 4.72 (14.9) & 7.87 (24.8)  \\
3 & 36.5 & 49.8 & 2.58 (8.14) & 7.74 (24.4) \\
4 & 25.8 & 35.2 & 1.82 (7.89) & 5.74 (24.9) \\
5 & 20.1 & 27.3 & 1.42 (4.48) & 8.05 (25.4) \\
10 & 9.63 & 13.0 & 0.680 (2.14) & 8.39 (26.5) \\
\hline
\end{tabular}
\end{ruledtabular}
\end{table}
\subsection{Insulating cylinder walls}

\begin{table}
\begin{ruledtabular}
\caption{\label{t4} The same as in Table \ref{t1} (wide gap, $\hat\eta=0.25$)
but for insulating cylinders.}
\begin{tabular}{lccccc}
\hline
$\hat\mu_B$ & Ha$_0$ & Ha$_1$ & $I_{\rm{axis}}$ [kA] &  $I_{\rm{fluid}}$ [kA]\\
\hline
-10 & 3.63 & 1.79 & 0.0421 (0.133) & -1.73 (-5.45) \\
-5  & 6.65 & 3.55 & 0.0836 (0.264) & -1.76 (-5.54) \\
-4  & 8.04 & 4.42 & 0.104 (0.328) & -1.77 (-5.58) \\
-3  & 10.3 & 5.89 & 0.139 (0.437) & -1.81 (-5.68) \\
-2  & 14.7 & 8.91 & 0.210 (0.662) & -1.89 (-5.60) \\
-1  & 30.6 & 20.5 & 0.483 (1.52) & -2.42 (-7.60) \\
\hline
{\bf 1} & $\infty$  &  {\bf 30.7} & {\bf 0.723 (2.28)} & {\bf 2.17 (6.84)} \\
2 & $\infty$    &  10.7 & 0.252 (0.794) & 1.76 (5.56) \\ 
3 & $\infty$    &  6.63 & 0.156 (0.492) & 1.72 (5.41) \\
4 & $\infty$    &  4.83 & 0.114 (0.359) & 1.71 (5.39) \\
5 &  17.4  & 3.81 & 0.0897 (0.283) & 1.70 (5.38) \\
10&  5.18  & 1.86 & 0.0438 (0.138) & 1.71 (5.38) \\
\hline
\end{tabular}
\end{ruledtabular}
\end{table}

\begin{table}
\begin{ruledtabular}
\caption{\label{t5} The same as in Table \ref{t2} (medium gap, $\hat\eta=0.5$)
but for insulating cylinders.}
\begin{tabular}{lccccc}
\hline
$\hat\mu_B$ & Ha$_0$ & Ha$_1$ & $I_{\rm{axis}}$ [kA] &  $I_{\rm{fluid}}$ [kA]\\
\hline
-10 & 6.09  &  4.66 &  0.190 (0.599) & -3.99 (-12.6) \\
-5  & 11.8  &  9.31 &  0.380( 1.20)  & -4.18 (-13.2) \\
-4  & 14.6  &  11.7 &  0.477 (1.50) & -4.29 (-13.5) \\
-3  & 19.4  &  15.7 &  0.640 (2.02) & -4.48 (-14.1) \\
-2  & 29.3  &  25.2 &  1.03  (3.24) & -5.15 (-16.2) \\
-1  & 73.6  &  64.6 &  2.63  (8.31) & -7.89 (-24.9) \\
\hline
{\bf 1}   & $\infty$  & {\bf 109.}  & {\bf 4.44  (14.0)} & {\bf 4.44 (14.0)} \\
2   & $\infty$  &  28.1  & 1.15  (3.61) & 3.45 (10.8) \\
3   & 32.6  &  17.2  & 0.701 (2.21) & 3.51 (11.1) \\
4   & 20.5  &  12.5  & 0.510 (1.61) & 3.57 (11.3) \\
5   & 15.3  &  9.81  & 0.400 (1.26) & 3.60 (11.3) \\
10  & 6.86  &  4.78  & 0.195 (0.615) & 3.71 (11.7) \\
\hline
\end{tabular}
\end{ruledtabular}
\end{table}

\begin{table}
\begin{ruledtabular}
\caption{\label{t6} The same as in Table \ref{t3} (narrow gap, $\hat\eta=0.75$)
but for insulating cylinders.}
\begin{tabular}{lccccc}
\hline
$\hat\mu_B$ & Ha$_0$ & Ha$_1$ & $I_{\rm{axis}}$ [kA] &  $I_{\rm{fluid}}$ [kA] \\
\hline
-10 & 14.2 & 13.8 & 0.975 (3.07) & -14.0 (-44.0) \\
-5  & 28.2 & 27.7 & 1.96 (6.17) & -15.0 (-47.3) \\
-4  & 35.4 & 34.9 & 2.46 (7.77) & -15.6 (-49.2) \\
-3  & 47.7 & 47.2 & 3.33 (10.5) & -16.7 (-52.5) \\
-2  & 74.8 & 74.7 & 5.28 (16.6) & -19.4 (-60.9) \\
-1  & 208. & 205. & 14.5 (45.7) & -33.8 (-107) \\
\hline
{\bf 1}   & $\infty$ & {\bf 464.} & {\bf 32.8 (103.)} & {\bf 10.9 (34.3)} \\
2   & 103. & 80.3 & 5.67 (17.9) & 9.45 (29.8) \\
3   & 56.2 & 49.1 & 3.47 (10.9) & 10.4 (32.7) \\
4   & 39.7 & 35.9 & 2.54 (7.99) & 11.0 (34.6) \\
5   & 30.9 & 28.4 & 2.18 (6.32) & 12.4 (35.8) \\
10  & 14.8 & 13.9 & 0.982 (3.10) & 12.1 (38.2) \\
\hline
\end{tabular}
\end{ruledtabular}
\end{table}

\begin{figure*}[htb]
\vbox{
\hbox{
\psfig{figure=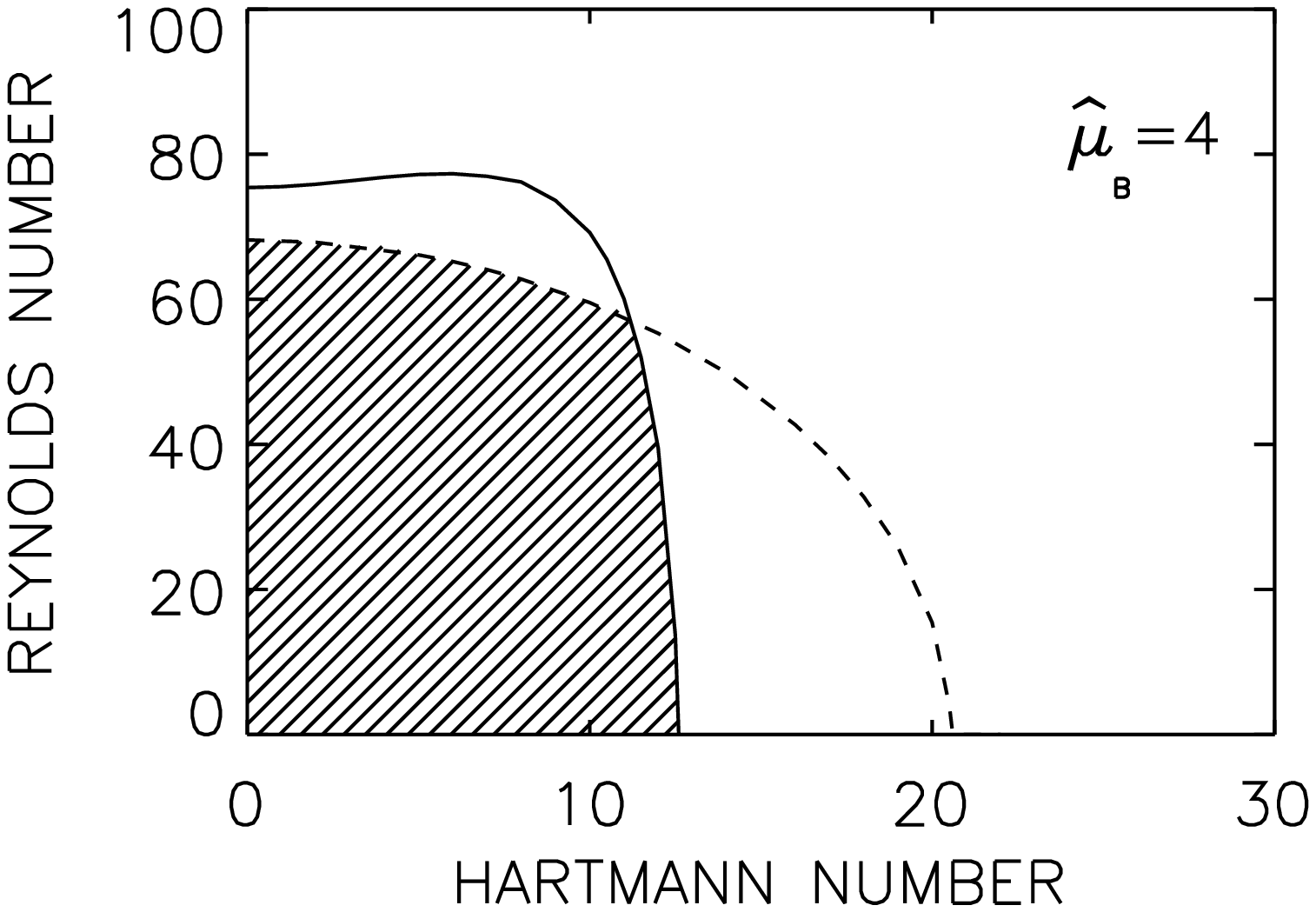,width=6cm,height=5cm}
\psfig{figure=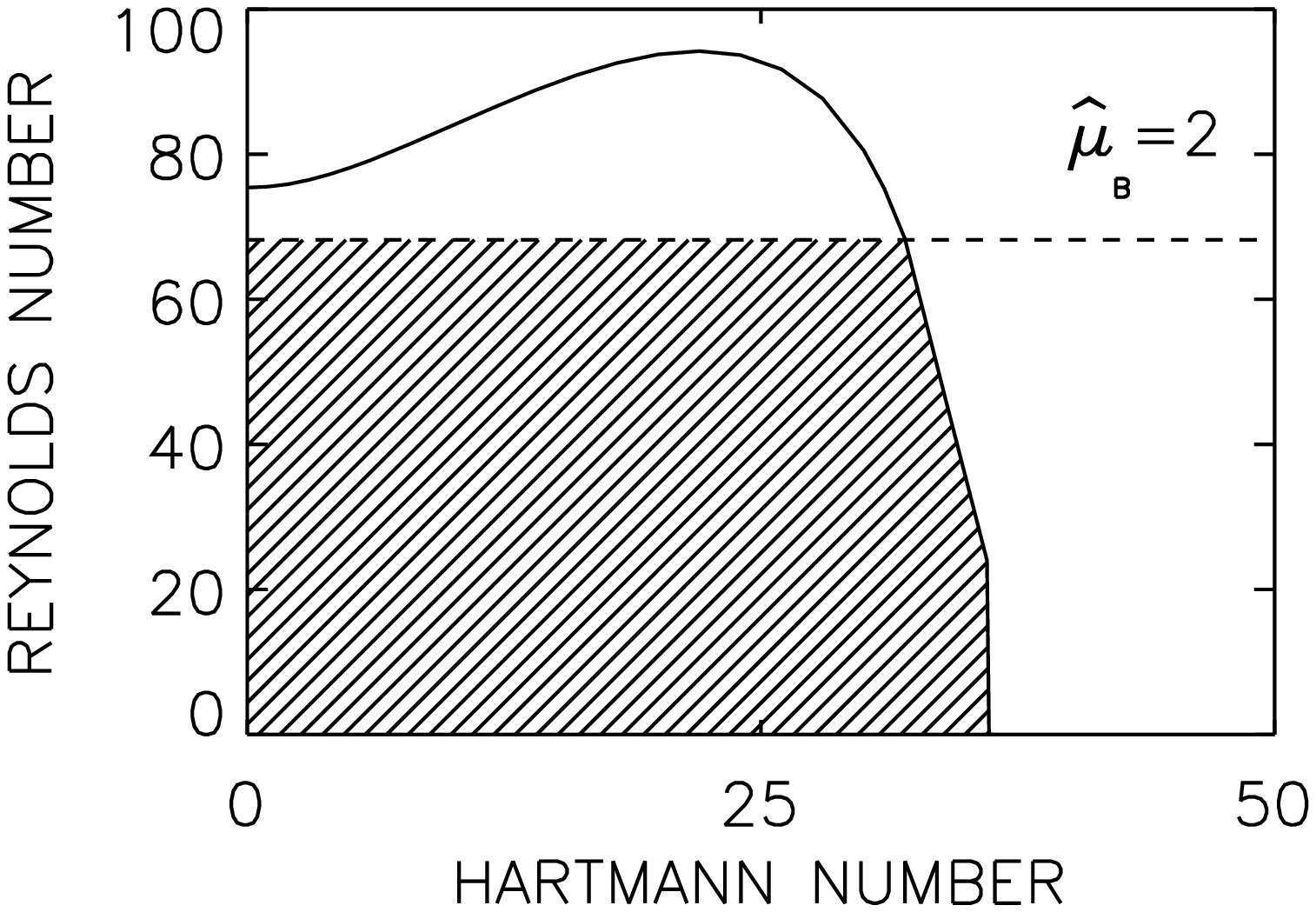,width=6cm,height=5cm}
\psfig{figure=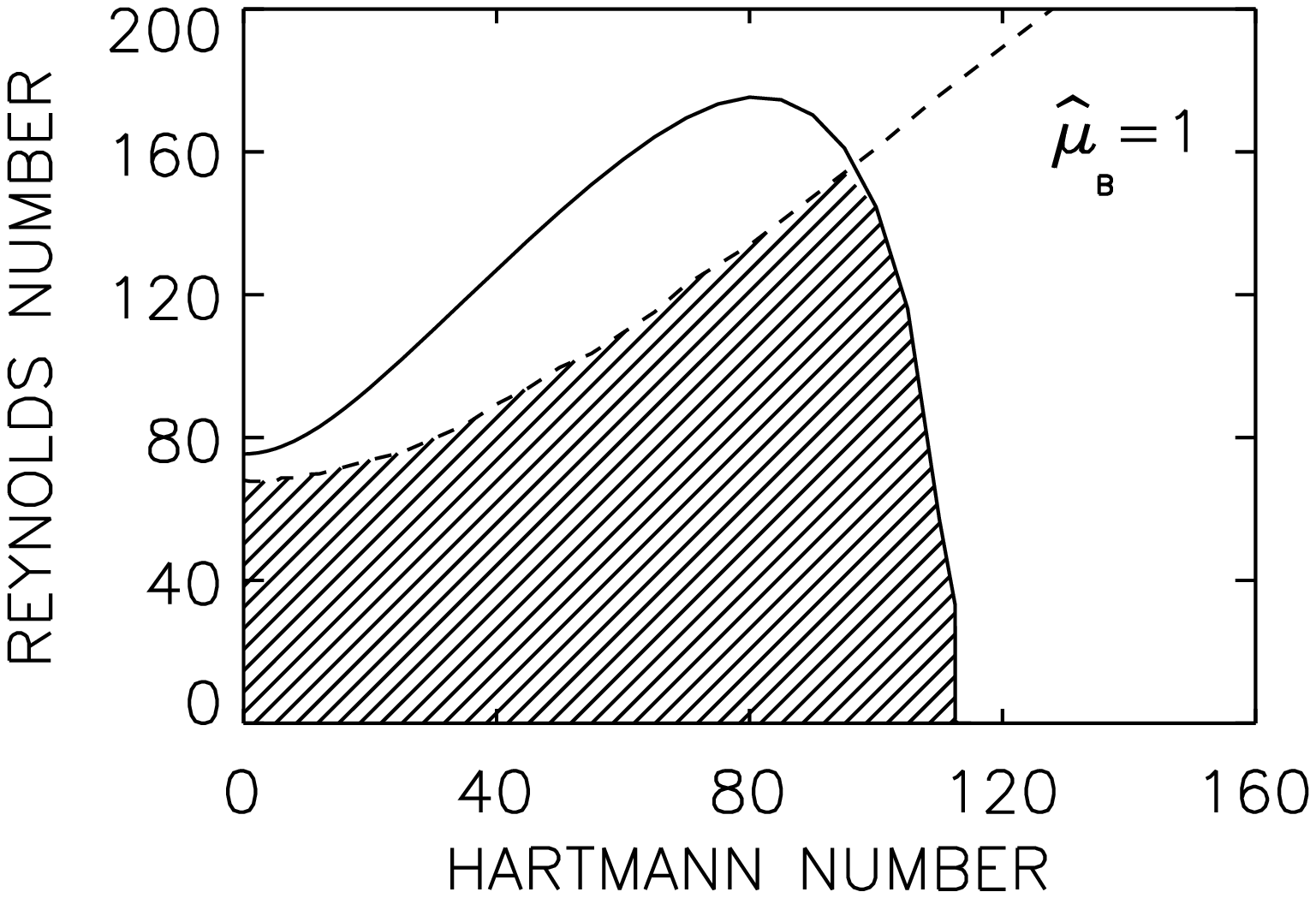,width=6cm,height=5cm}
}
\hbox{
\psfig{figure=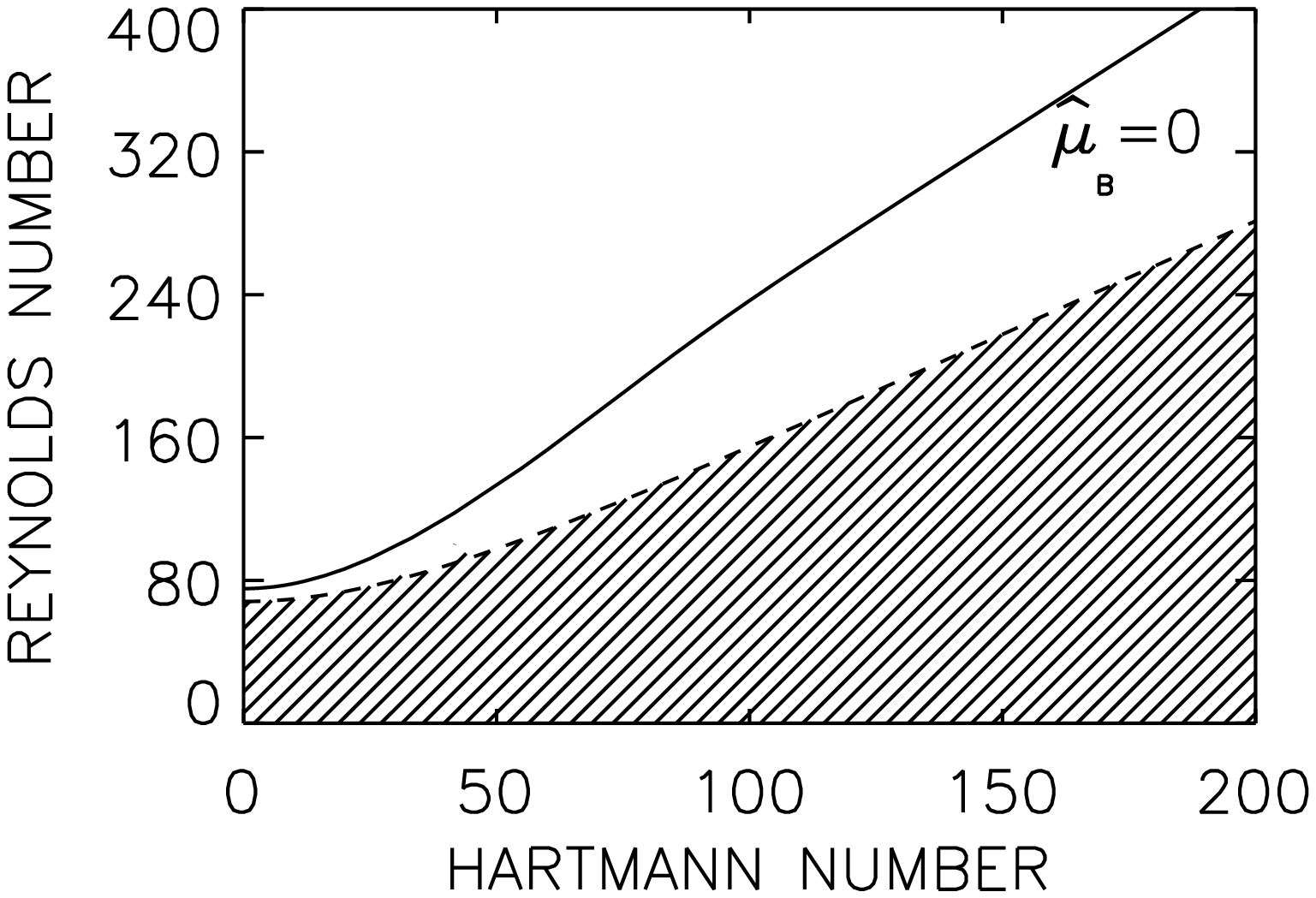,width=6cm,height=5cm}
\psfig{figure=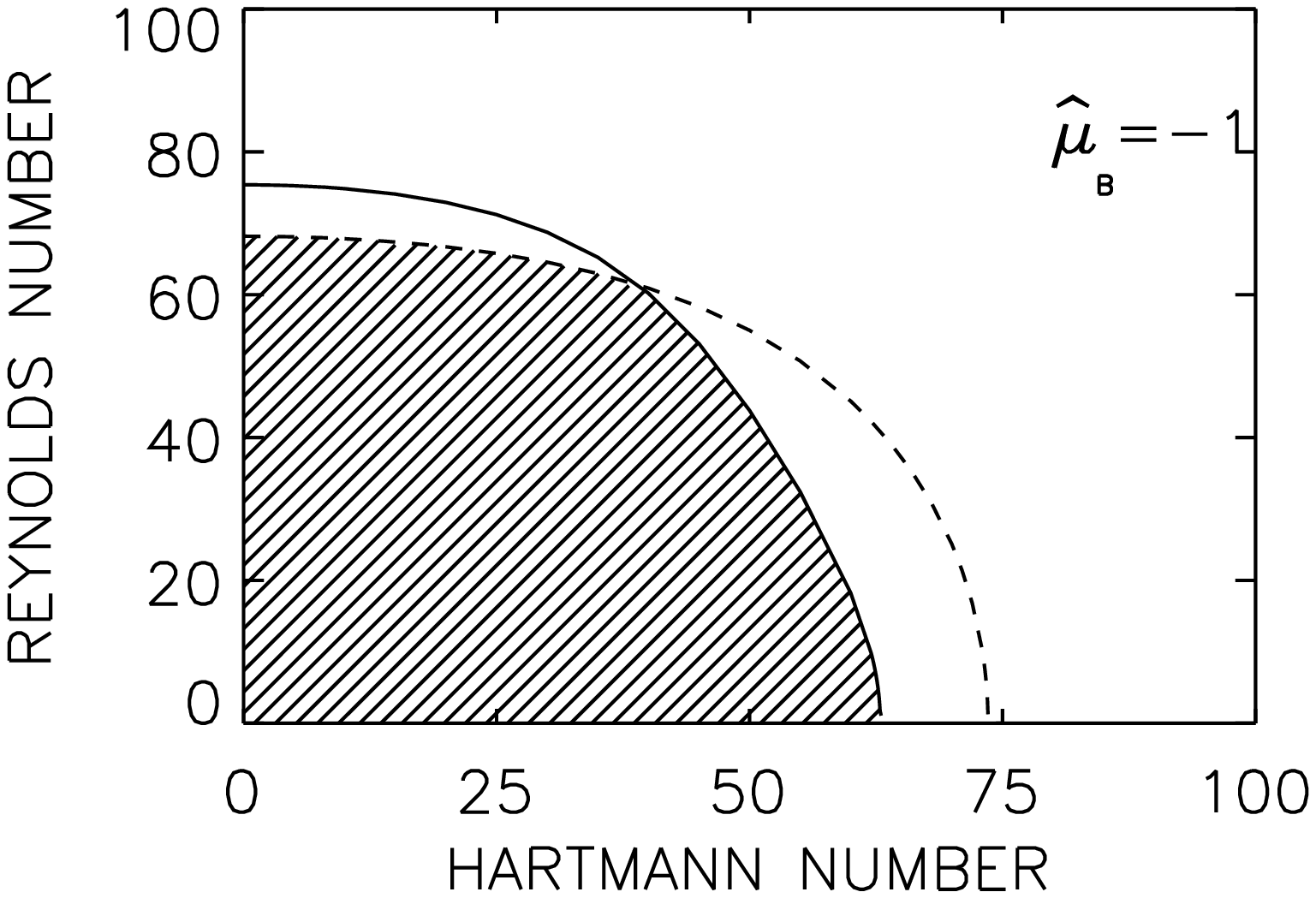,width=6cm,height=5cm}
\psfig{figure=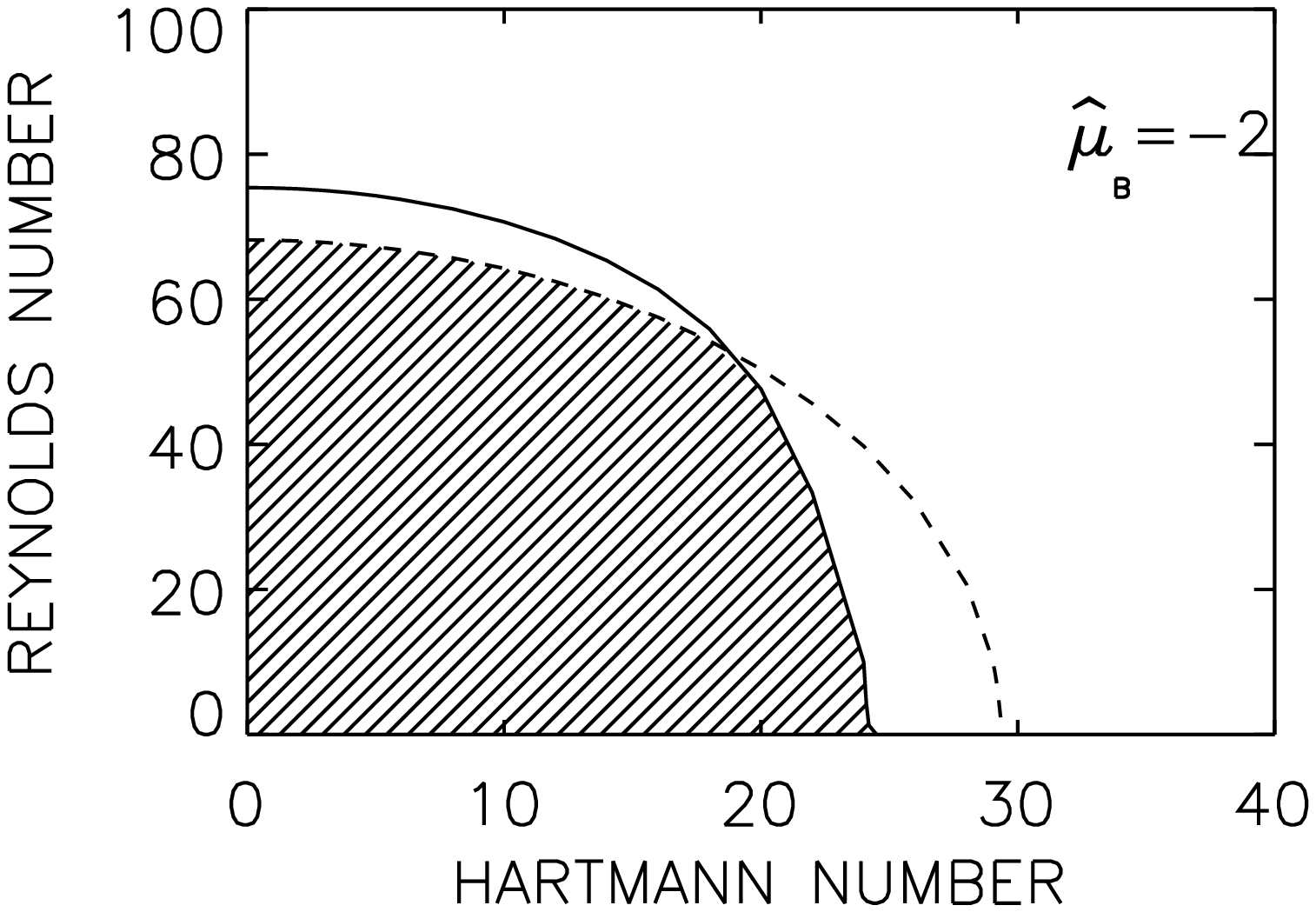,width=6cm,height=5cm}
}
}
\caption{\label{fig3} The same as in Fig. \ref{fig2} but for 
insulating cylinder walls.
}
\end{figure*}
Calculations were also done for insulating cylinder walls; the results
are given in Fig. \ref{fig3}.  They are generally similar as those for the
conducting cylinders, but with one important exception.  The $m=0$ and
$m=1$ stability curves now almost always cross each other, as they do for
conducting cylinders only for almost current-free $B_\phi$ profiles. One
can again observe for not too steep profiles (for $\hat\mu_B\simeq \pm 1$)
how for weak fields the $m=0$ mode {\em stabilizes} the rotation until
beyond the cross-over point the $m=1$ mode strongly {\em destabilizes} the
rotation.
\section{Conclusions}
We have shown how complex the interaction of magnetic fields and differential
rotation can be in Taylor-Couette flows, including also a strong dependence
on the magnetic Prandtl number.  For large Pm the field destabilizes the
differential rotation, whereas for small Pm it stabilizes it.  However, if
the field (or rather the current) is too great, then the Tayler instabilities
will always destabilize any differential rotation.

In order to prepare laboratory experiments, we also did calculations at
values of Pm appropriate for liquid metals, for both conducting and insulating
cylinder walls. In particular, we considered the almost uniform field profile
$\hat\mu_B=1$.  For both conducting and insulating boundaries, the field is
initially stabilizing, but after the most unstable mode switches from $m=0$
to $m=1$ it is strongly destabilizing, until the pure Tayler instability sets
in even at $\rm Re=0$.

For various gap widths and field profiles, we also computed the critical
Hartmann numbers and the corresponding electric currents. Tables II--VII
give the required currents for both conducting and insulating walls; note
how insulating walls (Tables V--VII) typically require lower currents
than conducting walls (Tables II--IV).  The other clear trend is that
the currents are lesser for wider gaps and greater for narrower gaps.
An optimal experiment might therefore have $\hat\eta=0.25$, insulating
walls, and $\hat\mu_B=1$, which would require only 6.84 kA even with
gallium-indium-tin (Table V).
\bigskip
\begin{acknowledgments}
The work of DS was supported by the Helmholtz Institute for Supercomputational
Physics in Potsdam and partly by the Leibniz Gemeinschaft under program SAW.
\end{acknowledgments}


\end{document}